\documentclass[12pt]{article}
\usepackage{epsfig,graphicx}

\begin{document}

\hfill Journ. de Physique, IV France {\bf {10}} (2000)
Pr3-183.

\vskip .6in

  {\Large{\bf{Topological defects in Spin Density Waves.}}}

\vskip .1in
\centerline{N. Kirova and S.Brazovskii.}
\vskip .1in
{\it 
{ Laboratoire de Physique Th\'eorique et des Mod\`eles Statistiques, CNRS,}}

{\it{B\^at.100, Universit\'e Paris-Sud, 91405 Orsay cedex, France.}}

{\it{ e-mail: kirova@ipno.in2p3.fr brazov@ipno.in2p3.fr }}
\vskip .2in

\begin{narrow}

{\bf Abstract.}

 The rich order parameter of Spin Density Waves allows
for an unusual object of a complex topological nature: a half-integer
dislocation combined with a semi-vortex of the staggered magnetization. It
becomes energetically preferable to ordinary dislocation due to enhanced
Coulomb interactions in the semiconducting regime. Generation of these
objects changes e.g. the narrow band noise frequency.
\end{narrow}

\vskip .2in

\section{ Introduction.}

Topological defects in Electronic Crystals - solitons, phase slips (PS) and
dislocation lines/loops (DLs) are ultimately necessary for the current
conversion and depining processes, see collections \cite{Charge,Electronic,
ECRYS-99}. Microscopically in Charge and Spin Density Waves (CDW, SDW, DW)
the PS starts as a self-trapping of electrons into solitons with their
subsequent aggregation (see \cite{BK,Braz-ch,Braz95} for review). Macroscopically
the PS develops as the edge DL proliferating/expanding across the sample 
\cite{Ong,Gorkov-ch}.

An important feature of semiconducting quasi one-dimensional DWs is the
Coulomb hardening \cite{Art} of their compressibility when the normal
carriers freeze out at low temperature $T$. Then the energetics of DL 
\cite{Braz91}, the accompanying electronic structure \cite{Kir99}, etc.  are
determined by the Coulomb forces limited by screening facilities of remnant
free carriers. 

The CDW/SDW are characterized by scalar/vector order parameters: $\eta
_{cdw}\sim \cos [Qx+\varphi ]$, $\vec{\eta}_{sdw}\sim \vec{m}\cos
[Qx+\varphi ]$ where $\vec{m}$ is the unit vector of the staggered
magnetization. Here we will show that SDWs allow for unusual $\pi $ PSs
forbidden in CDWs where only $2\pi $ PSs are allowed. Namely in SDW conventional dislocations loose their priority in favor of
special topological objects: a half-integer dislocation combined with a
semi-vortex of a staggered magnetization vector. 
Their possible manifestation may be found in a Narrow band Noise (NBN)
generation. The $\pi $- PSs reduce twice (down to its CDW value $\Omega
/j=\pi $) the universal ratio $\Omega /j$ of the fundamental NBN frequency $ 
\Omega $ to the mean sliding current $j$.  The splitting of the normal $2\pi $ 
-dislocation to the $\pi $ ones in energetically favorable due to Coulomb
interactions. The magnetic anisotropy
confines half-integer DLs in pairs connected by a magnetic domain wall.

Below we first consider the energy of usual DLs in DWs. Then we 
turn to the special case of  SDW and discuss combined topological
objects. Finally we discuss various models for the NBN generation and 
consequences of the existence of combined topological objects in SDW. 

\section{ A single dislocation in semiconducting density wave.}

Any stationary configuration is determined by minimization of the energy
functional 
\begin{equation}
W\{\varphi \}=\int \frac{d\vec{R}}{s}\frac{\hbar v_{F}}{4\pi }\left[
C_{\parallel }^{0}\left( \frac{\partial \varphi }{\partial x}\right)
^{2}+C_{\bot }(\nabla _{\bot }\varphi )^{2}\right] +W_{C}  \label{W}
\end{equation}
Here $v_{F}$ is the Fermi velocity of the parent metal, $\vec{R}=(x,\vec{r})$
with $x$ being the chain direction, $s=a_{\bot }^{2}$ is the area per 
chain. The Coulomb part of the energy, with some simplifications,
is 

\[
W_{C}=\int d\vec{R}_{1}d\vec{R}_{2}n_{c}(\vec{R}_{1})n_{c}(\vec{R}_{2})\frac{ 
\exp [-|\vec{R}_{1}-d\vec{R}_{2}|/r_{scr}]}{|\vec{R}_{1}-d\vec{R}_{2}|} 
\,,\;n_{c}=\frac{e\rho _{c}}{\pi s}\frac{\partial \varphi }{\partial x}
\]
where $n_{c}$ is the electric charge density of the locally deformed DW, $ 
\rho _{c}$ and $\rho _{n}=1-\rho _{c}$ are the normalized densities of the
condensate and of the normal carriers. Recall that $\rho _{n}\rightarrow 1$
at $T\rightarrow T_{c}^{0}$ and $\rho _{n}\sim \exp (-\Delta /T)$ at low $T$
being activated through the DW gap $2\Delta $. The parameters $C_{\Vert }^{0}
$ and $C_{\bot }$ are the compression and the share moduli; within our
normalization $C_{\Vert }^{0}=\rho _{c}$ and $C_{\bot }\sim \rho _{c}$. In
the Fourier representation 
\begin{equation}
W\{\varphi \}=\frac{\hbar v_{F}}{4\pi }\sum |\varphi _{k}|^{2}\left[ \rho
_{c}k_{\parallel }^{2}+C_{\bot }k_{\bot }^{2}+\frac{\rho
_{c}^{2}r_{0}^{-2}k_{\parallel }^{2}}{(k_{\parallel }^{2}+k_{\bot
}^{2}+r_{scr}^{-2})}\right]   \label{W-k}
\end{equation}
where $r_{0}=\sqrt{\hbar v_{F}s\epsilon /(8e^{2})}$ is the screening length
in the parent metal and $r_{scr}=r_{0}/\sqrt{\rho _{n}}$ is the actual
screening length in the DW. 

The Coulomb interactions drastically affect the charged phase deformations.
They may not be important yet only at shortest interchain distances $r<r_{0}$ 
, which are allowed only if $r_{0}>a_{\bot }$, where the usual elastic
theory is applicable. Beyond this core but still within the screening
distance $r_0<r<r_{scr}$ we can write the energy, collecting only the senior
terms, as 
\begin{equation}
W\{\varphi \}\approx \frac{\hbar v_{F}}{4\pi }\sum |\varphi _{k}|^{2}\left[
C_{\bot }k_{\bot }^{2}+r_{0}^{-2}\frac{k_{\parallel }^{2}}{k_{\bot }^{2}} 
\right] \,,\;k_{\bot }\gg r_{scr}  \label{unscr}
\end{equation}
This expression describes a nonanalitic elastic theory with energy dependent
on ratio of gradients rather on their values. Finally at large distances $ 
r\gg r_{scr}$ we have 
\begin{equation}
W\{\varphi \}\approx \frac{\hbar v_{F}}{4\pi }\sum |\varphi _{k}|^{2}\left[
C_{\parallel }k_{\parallel }^{2}+C_{\bot }k_{\bot }^{2}\right]
\,,\;C_{\parallel }=\rho _{c}/\rho _{n}\,,\;k_{\bot }\ll r_{scr}
\end{equation}
The effective elastic theory is restored but in stretched coordinates $(x 
\sqrt{C_{\bot }\rho _{n}/\rho _{c}},r)$. We summarize that the effective
compressibility $C_{\parallel }$ hardens with $r$ (starting from $ 
C_{\parallel }^{0}$ at shortest interchain distances) as $C_{\parallel }\sim
r^{2}/r_{0}^{2}$ beyond the screening length of the parent metal $ 
r>r_{0}\sim 1\AA $, until it saturates at $ 
r_{scr}$ at the value which grows activationally with $T$. 

Consider a D-loop of a radius $R$ in the $(y,z)$ plane embracing a number $ 
N=\pi R^{2}/s$ chains or a D-line stretched in $z$ direction at a distance $ 
Y=a_{\bot }N$ from its counterpart or from the surface. The DL is
characterized by its energy $W(N)$ and by the 'chemical potential' $ 
w=\partial W/\partial N$ which determines the mutual equilibrium among DLs
and between them and pairs of normal electrons. The scale of the DL energy
per unit length is $E_{0}/a_{\bot }$. Within the (effective) elastic theory
it is $E_{0}\sim \sqrt{C_{\parallel }C_{\bot }}\hbar v_{F}/a_{\bot }$ where $ 
C_{\bot }$ is a measure of the interchain coupling. For essentially quasi-1D
systems like typical CDWs, where the actual transition temperature $ 
T_{c}<T_{c}^{0}\sim \Delta $ (so that in the ordered phase always $\rho
_{c}\approx 1$) we have $E_{0}\sim T_{c}$. For systems with wider interchain
electronic bandwidth $t_{\bot }>\Delta $, which is typical for SDWs, the
scale is $ \tilde E_{0}\sim \rho _{c}t_{\bot }$ which can be larger than $ 
T_{c}\approx T_{c}^{0}$. A characteristic feature of the nonscreened regime 
(\ref{unscr}) is an optimal perpendicular scale $L_{\bot }=k_{\bot }^{-1}$ at
any given longitudinal scale $L_{\Vert }=k_{\Vert }^{-1}$: $L_{\bot
}^{2}\sim L_{\Vert }r_{0}C_{\bot }^{1/2}$. Now multiplying the energy
density $\sim C_{\bot }L_{\bot }^{-2}$ by the characteristic volume $ 
L_{\Vert }L_{\bot }^{2}$ we estimate the energy as a function of $R=L_{\bot }
$ as $\hbar v_{F}C_{\bot }^{1/2}R^{2}/r_{0}=w_{C}N$ with $N=\pi R^{2}/r_{0}$
and $w_{C}\sim E_{0}^0 a_{\bot }/r_{0}$, $E_{0}^0\sim \max [T_{c}, \rho_ct_{\bot
}]$.

Finally for the D-loop energy $ 
W_{D}(N)$ one finds \cite{Braz91}:\newline
1. $W_{D}(N)\sim \sqrt{N}\ln NE_{0}\ \ \ \ \ \ \qquad \qquad N\sim 1$. 
\newline
2.\rm{\ }$W_{D}(N)\sim \ Nw_{C}$, $w_{C}\sim E_{0}^{0}a_{\bot }/r_{0}\
\ \  \qquad R<r_{scr}\qquad (N=\pi R^{2}/s)\qquad \qquad \quad $ 
\newline
3. $W_{D}(N)\sim \sqrt{N}\ln NE_{0}r_{scr}/r_{0}\quad \qquad R>r_{scr}\qquad
\qquad \qquad \qquad \qquad \qquad \quad $\newline
For the dislocation energy within $r_{scr}$ there is not a usual perimetric
law $\sim \sqrt{N}$ but rather the area one $\sim N$. At large distances the
standard perimetric law is restored but with the greatly enhanced $\sim \rho
_{n}^{-1/2}$ magnitude. 

In SDW the energy of a pure magnetic vortex loop $W(\vec{m})$ is not
affected by the Coulomb forces, so that its only scale is ${E}_{0}$. We have 
\begin{equation}
W_{\vec{m}}\{\vec{m}\}=\int \frac{dxd\vec{r}}{s}\left[ \tilde{C}_{\parallel
}\left( \frac{\partial m}{\partial x}\right) ^{2}+\tilde{C}_{\bot }(\vec{ 
\nabla}_{\bot }\vec{m})^{2}\right] ,\qquad W_{\vec{m}}(N)\sim \sqrt{N}\ln N 
\tilde {E}_{0}
\end{equation}
where $\tilde{C}_{\parallel }$, $\tilde{C}_{\bot }$ are the elastic moduli
related to the rotation of the staggered magnetization unit vector $\vec{m}$ 
. They are similar to phase displacement moduli taken without Coulomb
interactions $C_{\parallel }^{0},\ C_{\bot }$, in (\ref{W}). Hence for
the VLs the regime 1. is applicable at all $N$.

The resulting $\mu _{D}$ of a single D-loop is drown schematically at Fig.1.
Here the dashed line corresponds to the magnetic vortex loop. The inner
region of the solid line describes both the D-loop and the magnetic vortex.

\section{ Half-integer dislocation combined with semi-vortex.}

In SDWs the Coulomb enhancement of the dislocation energy plays a principal
role to bring to life a special combined topological object. This is the
half-integer dislocation accompanied by the $180^{o}$ rotation $\mathcal{O} 
_{\pi }$ of the staggered magnetization $\vec{m}$. Indeed, the SDW order
parameter $\vec{\eta}=\vec{m}\cos (Qx+\varphi )$ allows for the following
three types of self-mapping $\vec{\eta}\rightarrow \vec{\eta}$.  (The mapping
is a general requirement of topological connectivity which selects the
allowed configurations \cite{Mermin, Mineev}.) \newline
i. normal dislocation: $\ \ \varphi \rightarrow \varphi +2\pi ,\ \vec{m} 
\rightarrow \vec{m}$;\newline
ii. normal $\vec{m}$ -- vortex: $\vec{m}\rightarrow \mathcal{O}_{2\pi }\vec{m 
},\ \ \ \varphi \rightarrow \varphi $;\newline
iii. combined object : $\ \ \varphi \rightarrow \varphi +\pi ,\ \ \vec{m} 
\rightarrow \mathcal{O}_{\pi }\vec{m}=-\vec{m}$.\newline
In the last case both the orientational factor $\vec{m}$ and the
translational one $\cos (Qx+\varphi )$ change the sign, but their product $ 
\vec{\eta}$ stays invariant. A necessity of semi-vortices in conventional
antiferromagnets in presence of frozen-in host lattice dislocations has been
realized already in \cite{ Dzyal}. In the SDW the semi-vortices become the
objects of the lowest energy created in the course of PS process. Indeed
only not far below $T_{c}$ at $\rho _{n}\sim 1$ the elastic moduli related
to the phase displacements and to magnetization rotations are of the same
order, hence all three objects have similar energies, $W_{D}\sim W_{\vec{m}}$ 
. With lowering $T$ the energy of the object ii. is not affected (except for
a universal dependence $\sim \rho _{c}$ near $T_{c}$) because charges are
not perturbed so that Coulomb forces are not involved. For objects i. and
iii. the major energy $\sim \rho _{n}^{-1/2}$ is associated to distortions
of $\varphi $ so that the energy of $\vec{m}$ rotation in the case iii. may
be neglected. To compare main contributions to energies of objects i. and
iii. we remind that at given $N$ the DL energy depends on its winding number 
$\nu _{\varphi }$ as $W_{D}\sim E_{0}(\rho _{s}/\rho _{n})\nu _{\varphi }^{2}
$, where $\nu _{\varphi }=1/2$ for $\pi $-DL and $\nu _{\varphi }=1$ for the 
$2\pi $-DL. The energy of the magnetic vortex depends on the vorticity $\nu
_{\vec{m}}$ as $W_{\vec{m}}\sim E_{0}\rho _{s}\nu _{\vec{m}}^{2}$. We must
compare their energies at the given number of accumulated electrons $2N$.
For shortness consider only the largest (screened) sizes of DLs.

For the D-loop with the radius $R$ and the D-line located at a distance $Y$
from its (spatially image) counterpart we have correspondingly\newline
$N\sim \nu R^{2},\qquad \mu _{D}\sim \partial (\nu _{\varphi }^{2}R\ln
R)/\partial (\nu _{\varphi }R^{2})\sim \nu _{\varphi }^{3/2}/N^{1/2}$\newline
$N\sim \nu Y,\qquad \mu _{D}\sim \partial (\nu _{\varphi }^{2}\ln
Y)/\partial (\nu _{\varphi }Y)\quad \sim \ \nu _{\varphi }^{2}/N\qquad $ 
\newline
In both cases the lowest energy per electron $W$ is given by an object with
smallest $\nu _{\varphi }=1/2$ i.e. by the combined one, $W=(W_{D}+W_{\vec{m} 
})/2\sim W_{D}/2$.
We conclude that in SDW the normal dislocation must split into two objects
of the combined topology with the repulsion between them. Apparently they
will have the same sign of the displacive half-integer winding numbers (the
total charge $2eN$ must be preserved). But the half-integer spin rotation
numbers should have opposite signs (to avoid a divergence of the magnetic
energy). In Fig.2 we present the vector field of the local SDW magnetization 
$\vec{\eta}$ for a single ``chimera''. The chain axis is horizontal. 

\begin{figure}[tbh]
\begin{minipage}[c]{.45\linewidth}
\includegraphics*[width=5.5cm]{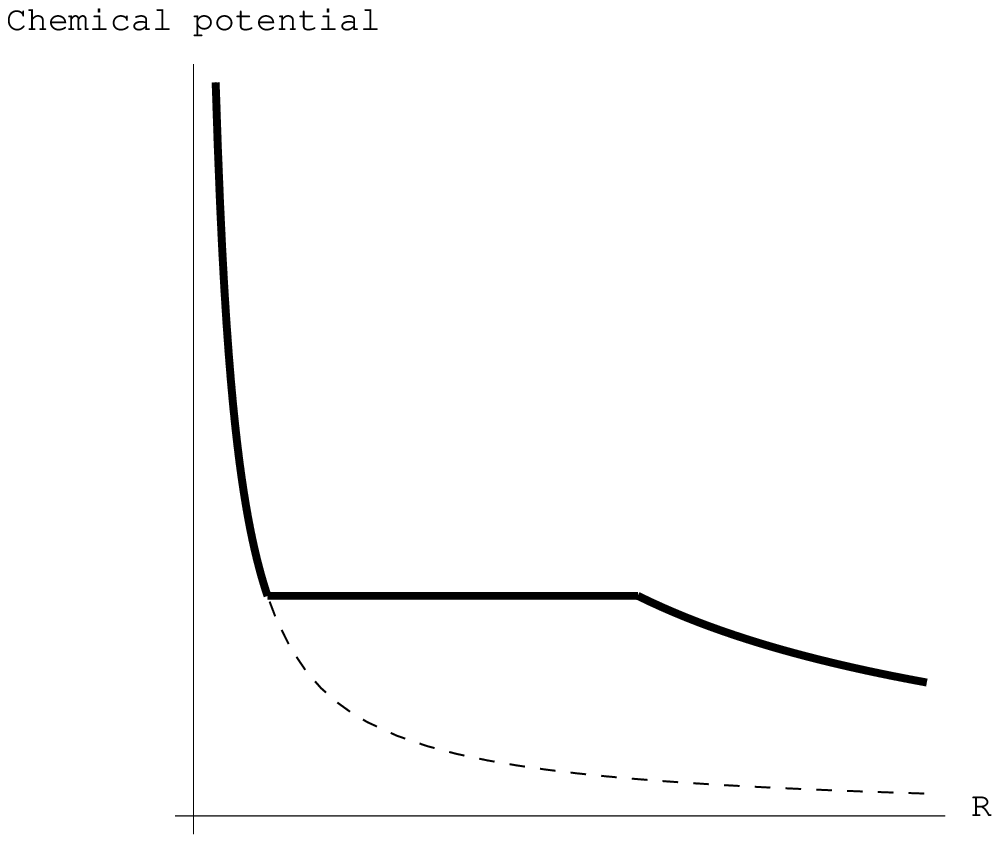}
\epsfxsize .7\hsize
\caption{The chemical potential $w=\partial W/\partial N$ of the D-loop. Dashed line corresponds to the model without Coulomb interactions and to the vortex loop.}
\end{minipage}\hfill 
\begin{minipage}[c]{.45\linewidth}
\includegraphics*[width=5.5cm]{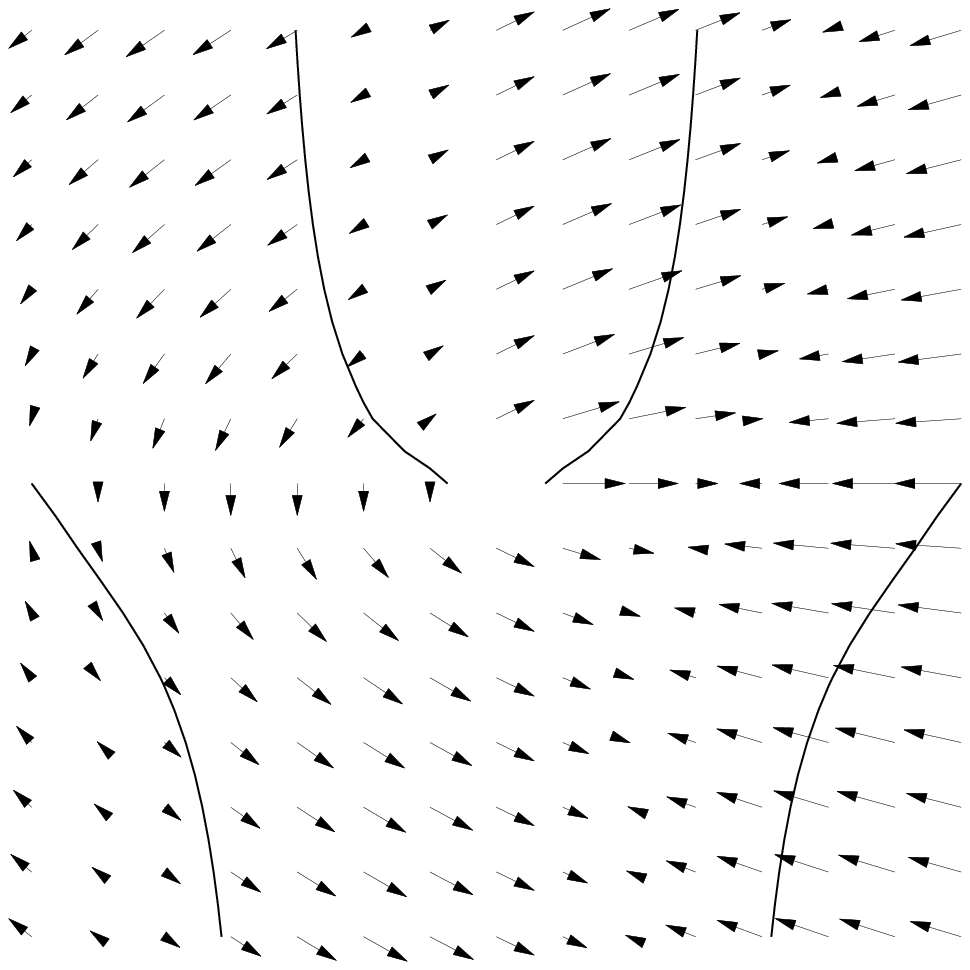}
\epsfxsize .7\hsize
\caption{Vector-field $\vec\eta$ for half-DL combined with semi-vortex.  Solid lines indicate  constant phases around the half-DL.}
\end{minipage}\hfill
\end{figure}

Consider now effects of a \emph{spin anisotropy} which have either
spin-orbital or dipole-dipole origin. The ``easy plane'' case allows for a
free rotation of spins, so it will not affect any of above conclusions. The
same will hold in case of a pure ``easy axis'' case but only at presence of
magnetic field $H>H_{sf}$ exceeding the spin-flop field $H_{sf}\sim 1T$
above which the spins will be tilted thus possessing a free rotation at the
hard plane. But the known SDW crystals have low symmetry, which originates
the spin anisotropy in all three directions. Being small, the anisotropy
will not affect the arrangement in a vicinity of the DL where the gradient
and the Coulomb energies dominate. But at large distances from the DL the
free rotation of spins is prohibited. The $\pi $-rotation of spins will be
concentrated in space within the Ne\'{e}l domain wall. It will form a string
(a plane in 3D) which confines the two combined objects. Indeed at $r<r_{scr}
$ the total energy gain with respect to the normal DL is $-E_{C}N/2$ while
the energy lost due to domain wall formation is $W_{{\vec{m}}}^{A}=w^{A}N,$
thus both having a similar $N$ dependence. Usually $w^{A}\sim 1K/chain<E_{C}$
and we have a constant repulsion between the two chimeras. But beyond the
screening volume $r>r_{scr}$ the total energy gain of two objects with
respect to one DL is $W=-(E_{0}/\sqrt{\rho _{n}})\ln N+w^{A}N$ - the Coulomb
energy slows down while the $W_{{\vec{m}}}^{A}$ keeps growing linearly.
Hence there is an equilibrium distance between the chimeras $N_{eq}\sim
E_{0}/(w^{A}\sqrt{\rho _{n}})$. The confined objects can be shown to split
off along the interchain direction. Usually the spin anisotropy is
noticeable only for one orientation and characterized by the spin-flop field 
$H_{s-f}\sim 1T$. It would originate the string of the length $\sim 0.1\mu m$ 
. At higher magnetic fields only a small in-plane anisotropy is left so that
the string length may reach the sample width which is typically $\sim 1\mu m$ 
. 

\section{Combined topological defects and the NBN generation.}

The DWs generate the Narrow Band Noise which is a coherent periodic
unharmonic signal with the fundamental frequency $\Omega $ being
proportional to the mean $dc$ sliding current $j$ with the universal ratio $ 
\Omega /j$ \cite{Richard}. In CDW ideally $\Omega /j=\pi $ which corresponds to carrying of
two electrons by displacing of the CDW by its wave length $\lambda $. In SDW
the ratio has been accessed only indirectly \cite{Jerome,Clark} with the
different experiments being in favor of either the $\Omega /j=\pi $ of the
twice higher ratio $\Omega /j=2\pi $ . Surprisingly the origin of such a
bright effect as the NBN has not been identified yet with main pictures
competing:\newline
\emph{The Wash-Board Frequency} (WBF) model suggests that the NBN is
generated extrinsically while the DW modulated charge passes through the
host lattice sites or its defects \cite{Zettle}.\newline
\emph{The Phase Slip Generation} (PSG) model suggests that the NBN is
generated by the phase discontinuities occurring near injecting contacts 
\cite{Ong,Gorkov-ch}.
Recall that CDW and SDW order parameters are 
\[
\eta_{cdw}\sim\cos[Qx+\varphi] \ \mathrm{and} \ \vec\eta_{sdw}\sim\vec m\cos[ 
Qx+\varphi]
\]
where $\vec m$ is the vector of the staggered magnetization. Importantly the
oscillating densities are 
\[
\rho_{cdw}\sim\eta_{cdw}\sim\cos[Qx+\varphi],\ \mathrm{but} \
\rho_{sdw}\sim\delta(\vec\eta_{cdw})^2\sim\cos[2Qx+2\varphi]
\]
So the SDW charge modulation wave length is only half of the CDW one $ 
\lambda=2\pi/Q$. Hence within the WBF model the NBN frequency $\Omega$ is
doubled: In CDW $\Omega=-\dot\varphi =\pi j$ while in SDW $ 
\Omega=-2\dot\varphi =2\pi j$.

At first sight the observation of the twice different ratios is natural from
the point of view of the WBF model. Nevertheless the WBF model has an
unresolved weaknesses. The original concept implied the interaction between
the rigid DW and the regular host lattice $\sim \cos (n\varphi )$ where the
commensurability index is typically $n=4$ which would give an $n$- fold WBF
contrary to experiments. Nowadays a common belief is that the necessary
potential $V_{imp}\sim \cos \varphi $ is provided by the host impurities.
But actually $V_{imp}\sim \cos (Qx_{i}+\varphi )$ so that the positionally
random phase shifts $\sim Qx_{i}$ will prevent any coherence in linear
response. (The mode locking (Shapiro steps) is still possible as a second
order nonlinear effect.) But for the linear effects the only possibility
left is to suppose that the DW does not slide at the sample surface so that
the coupling $\sim \cos (\varphi _{bulk}-\varphi _{surface})$ would provide
a necessary WBF.

The PSG model (see \cite{Gorkov-ch} and refs. in \cite{Kir99,Req}) is
attractive because the PSs are ultimately necessary to provide the current
conversion at the contacts. A weak point of the PSG model is to explain
their regularity: e.g. only one DL can flash across at a given time while
the next DW is waiting for the next DW period to pass. The topological
connectivity requires that after the phase slip or going around the
dislocation the order parameter should be mapped onto itself. For crystals
it means that exactly one lattice period enters or in terms of dislocation
language one says that the dislocation Burgers vector coincides with the
lattice period, which is $2\pi $ in $x$ - direction in our case. Since the
unit sells of the CDW and the SDW are the same (2 electrons) the $\Omega
/j=\pi $ would be the same, which contradicts to the latest experimental
results \cite{Clark}. This argument has been used to exclude the PSG
mechanism in favor to the WBF one. A generation of half-integer DLs resolves
this contradiction by twice increase of $\Omega /j$. These ``chimeras'' are
energetically favorable at low temperatures, when the number of free
carriers is small. But near the SDW transition temperature the energies of a
magnetic vortex and of a the dislocation are comparable. Then the
possibility of usual DL to split into two combined objects depends on the
material parameters. It means that for SDW near $T_{c}$ the ratio $\Omega /j$
in not universal. For some materials it can be $2\pi $ in the whole
temperature range if the splitting of dislocation is energetically favorable
already near $T_{c}$, for other materials the ratio $\Omega /j$ can change
from $\Omega /j=\pi $ at high temperatures to $\Omega /j=2\pi $ at low
temperatures passing through the region of intermediate values when both
simple dislocations and ``chimeras'' coexist. Notice also a necessity of
studying the narrow sample which is indeed the case for molecular crystal
with SDWs. Otherwise the string length created by the spin anisotropy will
be shorter than the sample width and the chimers will propagate in loosely
bound pairs.

\section{ Conclusions.}

We conclude that the sliding SDW should generate ''chimers'': the combined
topological objects where the spin rotations are coupled to the DW
displacements. They are stable by lowering the DL Coulomb energy. This
combination effectively reduces the SDW period allowing e.g. for the twice
increase in the NBN frequency, which is an important disputable question.
The interest in such unusual topological objects may go far beyond the NBN
generation or the current conversion problem in SDWs. Actually the studies
on these complex patterns would correlate to current interest in formation
of topological objects from superfluid $^{3}He$ to the model of earlier
Universe \cite{LesHouches}.

\end{document}